\documentclass[11pt,twoside]{article}
\usepackage{newpasp666}
\usepackage{epsf}

\index{Gibson, B. K.}
\index{Giroux, M. L.}
\index{Shull, J. M.}
\index{Stocke, J. T.}

\begin{document}

\title{Do High-Velocity Clouds {\it Really} Fuel Galactic Star Formation?}
\author{Brad K. Gibson, Mark L. Giroux, John T. Stocke, J. Michael Shull}
\affil{Center for Astrophysics \& Space Astronomy, University of Colorado,
Boulder, CO 80309-0389}

\begin{abstract}
Tantalizing evidence has been presented supporting the suggestion that
a large population of extragalactic gas clouds permeates the Local Group, a
population which has been associated with the Galactic High-Velocity
Clouds (HVCs).  We comment on both the strengths and weaknesses
of this suggestion, informally referred to
as the Blitz/Spergel picture.
Theoretical predictions for the spatial and kinematic distributions, 
metallicities, distances, and emission properties of Blitz/Spergel
HVCs will be confronted with extant observational data.  
\end{abstract}


\section{Introduction}

Simulations of the
Local Group's formation (Klypin et~al. 1999) predict that an order of
magnitude more satellites should be associated with the Milky Way and M31 than
are actually observed.  This discrepancy is a significant challenge to
hierarchical clustering scenarios.  An intriguing suggestion as to the
whereabouts of the ``missing'' satellites is provided by the Local Group infall
model of Blitz et~al. (1999), who speculate that a large fraction of the
classical ensemble of HVCs are these Local Group building blocks.  The
continuing infall of the HVCs onto the disk of the Galaxy would then provide
the bulk of the fuel necessary to maintain ongoing star formation.

HVCs are ubiquitous ($\sim$20\% sky covering fraction)
clouds seen in HI emission, whose velocities are incompatible with
simple models of Galactic rotation.  Because the majority of
their distances are effectively
unconstrained, rampant speculation exists as to their exact nature and origin,
ranging from solar metallicity Galactic fountain gas ($d\la 10$\,kpc and
Z$\sim$Z$_\odot$), to Magellanic Cloud tidal debris ($d\la 50$\,kpc and
Z$\ga$0.25\,Z$_\odot$), to the Blitz/Spergel Local Group formation remnants
($d\ga 400$\,kpc and 0.0$\la$Z$\la$0.1\,Z$_\odot$).

The fact that each scenario makes specific predictions regarding the 
distance and metallicity for the ``typical'' HVC means that, in principle,
the above models could be distinguished from one another wth appropriate
observations.  In practice, 
the definitive observational discriminant has been difficult to obtain.

\section{Distances}

The cleanest discriminant between the competing HVC models is that of their
distance.  If it could be shown that the majority of HVCs reside in the
Galactic halo, as opposed to being distributed throughout the Local Group, one
could sound the death knell for the Blitz/Spergel model.  Unfortunately,
direct distance determinations for HVCs are few and far between;
to set a useful upper limit requires a suitably bright 
background halo star of known distance to lie directly behind a
high HI column density HVC.  The dearth of catalogued
blue horizontal branch stars and early subdwarfs in the outer halo (RR~Lyrae 
stars 
can sometimes be employed, in a pinch) is one immediate problem; those 
bright enough to obtain high S/N, high-resolution spectra (to
actually search for HVC absorption features) are rarer still.  Non-detections
(both for foreground and background probes) are more difficult to
interpret, as fine-scale HI structure may conspire to make the probe ``miss''
any intervening HI.

To date, there are only five HVCs for which either an upper limit or distance
bracket exists.  As Table~1 shows, of these five HVCs none is consistent with
an intra-Local Group residence, as might be expected under the Blitz/Spergel
picture.  An ongoing attempt to detect Complex~WD in absorption towards a
distant halo RR~Lyrae star may soon add a sixth entry to Table~1 (Comeron
2000).
A few other HVCs have solid lower distance limits, but they do not
provide any discriminant between halo and Local Group residence (being
only $\ga$1$-$5\,kpc).  These are therefore not reported here.

\begin{table}[ht]
\begin{center}
\caption{HVC Distances: Upper Limits or Distance Brackets}
\begin{tabular}{lcl}
\tableline
   HVC         & Distance (kpc) & Reference \\
\tableline
100$-$$\,\,\,$7$+$110 & $<$1           & Bates et~al. (1991)         \\
Complex~M             & $<$4           & Ryans et~al. (1997)         \\
328$-$16$+$100        & $\;\,<$11      & Sembach et~al. (1991)       \\
Complex~A             & 4$-$10         & van~Woerden et~al. (1999)   \\
279$-$33$+$120        & $\;\,<$50      & Richter et~al. (1999)       \\
\tableline
\tableline
\end{tabular}
\end{center}
\end{table}

The background stellar probe technique described above is virtually impossible
to apply to any potential
Local Group HVC at $d\ga$400\,kpc.  Perhaps the most promising
method for attempting to prove an HVC truly lies at $\sim$Mpc distances lies
in the detection of the tip of the red giant branch in any putative stellar
population associated with the HVC (Grebel et~al. 2000).

Recently, Combes \& Charmandaris (2000) have shown that both the Wakker \&
Schwarz (1991) and Braun \& Burton (1999) HVCs (at 1$^\prime$ and 30$^\prime$
resolution, respectively) follow closely the size-linewidth relation defined by
Galactic molecular clouds, \it provided that
their mean distances are $\sim$20\,kpc.
\rm  This is \it indirect \rm evidence against the Blitz/Spergel picture,
but concerns regarding the use of the size-linewidth technique as a distance
determinator must be heeded (Wakker \& van~Woerden 1997; \S~4.1).

\section{Kinematics and Spatial Deployment}

Both Blitz et~al. (1999) and Braun \& Burton (1999) have used the fact that
the dispersion $\sigma_{\rm LSR}$ 
in the HVC distribution relative to $v_{\rm LSR}$
is greater than the dispersion $\sigma_{\rm GSR}$
relative to $v_{\rm GSR}$ or $v_{\rm LGSR}$
as support for preferring the Galactic and Local Group standards of rest, over
the local standard of rest.  They use this as indirect support for an 
extragalactic origin for many HVCs.  

It should be stressed that, while $\sigma_{\rm GSR}<\sigma_{\rm LSR}$
\it is \rm a necessary condition for the Blitz/Spergel picture, it does
\it not \rm constitute sufficient proof.  Any
model that predicts a sinusoidal $v_{\rm LSR}$ vs. Galactic longitude
distribution, necessarily satisfies the same $\sigma_{\rm GSR}<\sigma_{\rm
LSR}$ condition, a wholly underappreciated fact.  Specifically, 
$\sigma_{\rm GSR}<\sigma_{\rm LSR}$ for \it all \rm Local Group infall \it and
\rm Galactic fountain \it and \rm Magellanic Stream disruption models.
In addition, there is a significant selection effect at play in these
$\sigma_{\rm GSR}$ vs $\sigma_{\rm LSR}$ comparisons in that 
$|v_{\rm LSR}|$$\la$100\,km\,s$^{-1}$ HI is not included in the
$\sigma_{\rm GSR}$ $\Rightarrow$ $\sigma_{\rm LSR}$ conversion.  Any
effect this ``missing'' gas might have upon the resulting
distribution was neglected by Blitz et~al. (1999) and Braun \& Burton (1999).

The superposition of Wakker's (1990; Ch.~5) Galactic fountain and Magellanic
Stream models results in an HVC flux distribution indistinguishable
from that observed.  Specifically,
sum Figures~9(b) and 9(d) of Ch.~5 in Wakker (1990)
and contrast with Figure~5 of Ch.~2.  Not only is the flux distribution
reproduced, but so are many secondary signatures, including the asymmetry in 
the velocity extrema (Blitz et~al. 1999; \S~5.2.3).  To be fair,
it should be noted that the spatial deployment of the HVCs (in the $\ell-b$
Aitoff projection) in Wakker's Galactic fountain $+$ Magellanic Stream model 
(top panels of Fig.~3 and 8 of Ch.~5 of Wakker 1990) appears inferior to 
that of the Blitz et~al. (1999; Fig.~13) model.  New Galactic fountain models
from de~Avillez (2000) may reduce this discrepancy.

\section{Metallicities}

High spectral resolution and high S/N ultraviolet spectra
taken with spectrometers
aboard HST and FUSE have enabled
major advances in quantifying the distribution of HVC
metallicities.  Preferred atomic lines for such abundance analyses include
those of SII and OI ($\alpha$-elements,
minimal dependency upon ionization corrections, relatively
insensitive to dust depletion);
other popular lines include those of FeII, SiII, and MgII (albeit
subject to large depletion corrections)
and NI (dominated by secondary production at low metallicities, however).

Regardless of atomic species considered, \it many \rm abundance determinations
are susceptible to uncertain dust depletion corrections, \it most \rm 
are susceptible to uncertain ionization corrections, and \it all \rm
are subject to (potentially considerable) spatial resolution
uncertainties.  The latter cannot be emphasized strongly enough.
The absorption line is probing sub-arcsecond scales
in the intervening HVC gas, while the 21cm HI is probing a scale 2$-$3 orders
of magnitude greater!  Substantial systematic uncertainties are typically
incorporated into the final quoted result to reflect this ``resolution''
uncertainty, but one can never be certain that a pathological case has not been
encountered.

Recall from \S~1 that in the Blitz/Spergel picture,
metallicities of the order $\la$0.1\,Z$_\odot$ are to be expected (but not
necessarily primordial), while
under a Galactic fountain scenario values of $\sim$Z$_\odot$ should be more
prevalent.  Magellanic Cloud debris might be expected to lie in the
$\sim$0.2$-$0.4\,Z$_\odot$ regime.  
Table~2 lists eight recent HVC metallicity
determinations; it is by no means complete, but it
does provide a representative sample.

\begin{table}[ht]
\begin{center}
\caption{High-Velocity Cloud Abundances}
\begin{tabular}{lcl}
\tableline
   HVC         & Abundance & Reference \\
\tableline
Complex~C             & 0.1$-$0.4\,S/H$_\odot$ & Wakker et~al. (1999);       \\
                      &                        & Gibson et~al. (2001a)       \\
Complex~WB            & $\ga$0.1\,Ca/H$_\odot$ & Robertson et~al. (1991)     \\
287$+$22$+$240        & 0.3\,S/H$_\odot$       & Lu et~al. (1998)            \\
Magellanic Stream     & 0.3\,S/H$_\odot$       & Gibson et~al. (2000)        \\
225$-$83$-$200        & $\ga$0.3\,Fe/H$_\odot$ & Gibson et~al. (2001b)       \\
279$-$33$+$120        & $\ga$0.5\,Fe/H$_\odot$ & Richter et~al. (1999)       \\
258$-$39$+$232        & $\ga$0.6\,Si/H$_\odot$ & Sahu \& Blades (197)        \\
100$-$$\,\,\,$7$+$110 & 0.7\,O/H$_\odot$       & Bates et~al. (1990,1991)    \\
\tableline
\tableline
\end{tabular}
\end{center}
\end{table}

An immediate conclusion to be drawn from Table~2 is that, \it to date\rm, no 
HVC shows unequivocal evidence for Z$<$0.1\,Z$_\odot$
(Blitz/Spergel).  Conversely, no 
HVC shows unequivocal evidence for Z$>$1.0\,Z$_\odot$ (Galactic
fountain), an interesting conundrum!  Future HST/STIS and FUSE analyses
are sorely needed, particularly in light of the fact that the anti-Local Group
barycentre clouds, where the Blitz/Spergel picture predicts an excess of
intra-Local Group HVCs, have yet to be sampled in UV absorption.

According to simple models of
Galactic chemical evolution with gas infall (Tosi 1988), Z$_{\rm inf}$ must be
$\la$0.1\,Z$_\odot$; for Z$_{\rm inf}>0.15$, the resulting stellar population
distribution violates the present-day Galactic disk constraints.  Table~2
would then imply that the majority of HVCs are not representative
of the class of infalling gas clouds invoked by theorists to explain the
G-dwarf problem (the status of Complex~C is still being debated at this time).

While not listed in Table~2, HVC~125$+$41$-$207 deserves special mention.
This HVC has received attention from Braun \& Burton (1999) owing to its
exceptionally narrow core 
HI linewidth ($\la$2\,km\,s$^{-1}$).  An indirect argument
based upon thermal and pressure characteristics suggests that the distance to
this HVC is of the order $\sim$700\,kpc (the quoted distance is based upon
an unpublished modification of the 
Wolfire et~al. 1995 model, and so a formal uncertainty in 
this number is not yet available).  Very low abundances have been claimed
for this HVC based upon MgII\,$\lambda\lambda$2796,2803 derived from the
HST/GHRS spectrum of Bowen \& Blades (1993) and Bowen et~al. (1995)
along the line of sight to the background AGN Mrk~205.
The former group found W$_\lambda$(MgII\,2796)=76$\pm$12\,m\AA\ and
W$_\lambda$(MgII\,2803)=46$\pm$17\,m\AA, 
while the latter found
W$_\lambda$(MgII\,2796)=169$\pm$28\,m\AA\ and
W$_\lambda$(MgII\,2803)=73$\pm$30\,m\AA, 
based upon the same spectrum; the source of the discrepancy is not stated.
Our conservative analysis of their dataset yields
W$_\lambda$(MgII\,2796)=130$\pm$40\,m\AA\ and
W$_\lambda$(MgII\,2803)=90$\pm$40\,m\AA.  A $b$-value of 6$\pm$1\,km\,s$^{-1}$
was found by all groups.  Constructing curves of growth, our analysis
implies a MgII abundance (68\% c.l.) of 
(0.03$-$0.19)\,Mg/H$_\odot$.  Dust depletion similar to that seen in the
Galactic halo clouds would increase this value by a factor of 3$-$4.
Unfortunately, the low S/N FUSE spectrum for Mrk~205 has (thus far) been
unable to shed new light on the metallicity of HVC~125$+$41$-$207, but a
scheduled HST/STIS programme (PID\#8625) should clarify the situation.


\section{The Magellanic Stream}

Special mention should be made of the most spectacular of HVCs, the Magellanic
Stream (MS), an $\sim$1000\,deg$^2$ HI feature trailing the Magellanic Clouds.
The MS is the one HVC for which we know the origin: the Clouds themselves.
What remains controversial is the exact mechanism by which this gas was
extracted from said Clouds: tides or ram pressure stripping?
The key discriminants between the two scenarios are: (i) the prediction of 
a leading counterpart to the trailing Stream (tidal models); and (ii) the tip
of the Stream (MS\,IV$\rightarrow$MS\,VI) being 2$-$3$\times$ further
away in the case of tides, than in the case of drag.  
A consequence of (ii) is the prediction that, in the
mean, the intensity of H$\alpha$ emission I(H$\alpha$)
from the Stream should decrease 
(increase) the further one moves down-Stream, under the tidal (drag) scenario
[under the inherent assumption that I(H$\alpha$) is driven primarily by
photo-ionization, as opposed to collisional ionization (c.f. Weiner
et~al. 2001)].

The extant observational data on the Stream favours the tidal
scenario, although there are admittedly one or two weak links:

\begin{itemize}
\item \vspace{-2.0mm} existence of Leading Arm Feature (LAF) 
$\Rightarrow$ tides 
(Putman et~al.  1998)
\item \vspace{-2.0mm} metallicity of Leading Arm 
HVC~287$+$22$+$240 $\Rightarrow$ tides (Lu et~al. 1998)
\item \vspace{-2.0mm} \it if \rm
I(H$\alpha$)$_{\rm MSV,VI}$$<$I(H$\alpha$)$_{\rm MSII}$ 
$\Rightarrow$ tides
\item \vspace{-2.0mm} I(H$\alpha$)$_{\rm LAF}$$<<$I(H$\alpha$)$_{\rm MS}$ 
$\Rightarrow$ not tides (Bland-Hawthorn et~al. 2001)
\item \vspace{-2.0mm} stars not observed in MS $\Rightarrow$ tides \it or \rm
drag (Yoshizawa 1998; Mihos 2001)
\item \vspace{-2.0mm} star streams in vicinity of LMC $\Rightarrow$ tides 
(Majewski et~al.  1999)
\item \vspace{-2.0mm} \it if \rm 
N$_{\rm HII,disk}$(65~kpc)$\la$10$^{19}$\,cm$^{-2}$ \it or if \rm 
n$_{\rm halo}$(50\,kpc)$\neq$(5$\pm$1)$\times$10$^{-5}$\,cm$^{-3}$\hfil\break
$\Rightarrow$ not Moore \& Davis (1994) drag
\end{itemize}

\section{Conclusions}

Perhaps the most important conclusion to take from the study of HVCs is that
they are \it not \rm all the same beasts.  To categorize all HVCs as
due to a fountain (Galactic waste) or infalling remnants of the Local Group's
formation (Galactic fuel) is doomed to failure from the start.  No doubt there
are a plethora of origins scattered about the HVC family tree.  
What can be concluded is that (to date) there is no unequivocal
evidence for the existence of intra-Local Group HVCs, based upon limited
distance, metallicity, and optical emissivity arguments.  Concerning the
latter, neither Weiner et~al. (2001) nor Tufte et~al. (2001) have thus far
observed an HVC consistent with the I(H$\alpha$) predictions of the 
Blitz/Spergel model (c.f. Bland-Hawthorn et~al. 2001).  On the other hand,
neither is
there unequivocal evidence for the existence of Z$\ga$Z$_\odot$ HVCs, as might 
be expected for a Galactic fountain origin (intermediate-velocity clouds
on the other hand do
occasionally show Z$\sim$Z$_\odot$, but that is a story left for another day).

\acknowledgments

We acknowledge the financial support at the University of Colorado
of the NASA LTSA Program (NAG5-7262), the
FUSE Science Team (NAS5-32985), and the FUSE GI Program (NAG5-9018).
A special thanks to all those who
helped BKG and JTS
consume excess quantities of excellent pisco sours, cigars, and sub-standard 
scotch -- Joss Bland-Hawthorn, Phil Maloney, Sylvain Veilleux, 
Kevin McLin, D.~J. Pisano, Ben Weiner, and Todd Tripp.


\begin{references}
{\small
\reference Bates, B, Catney, M.G., Keenan, F.P. 1990, \mnras, 242, 267
\reference Bates, B, Catney, M.G., Gilheany, S., et~al. 1991, \mnras, 249, 282
\reference Bland-Hawthorn, J., et~al. 2001, these proceedings
\reference Blitz, L., Spergel, D.N., Teuben, P., Hartmann, D. \& Burton, W.B.
1999, \apj, 514, 818
\reference Bowen, D.V. \& Blades, J.C. 1993, \apj, 403, L55
\reference Bowen, D.V., Blades, J.C. \& Pettini, M. 1995, \apj, 448, 662
\reference Braun, R. \& Burton, W.B. 1999, \aap, 341, 437
\reference Combes, F. \& Charmandaris, V. 2000, \aap, 357, 75
\reference Comeron, F. 2000, private communication
\reference de~Avillez, M.A. 2000, preprint (astro-ph/0001297)
\reference Gibson, B.K., Giroux, M.L., Penton, S.V.,
et~al. 2000, \aj, in press (astro-ph/0007078)
\reference Gibson, B.K., Giroux, M.L., Penton, S.V., et~al. 2001a, \aj, 
submitted
\reference Gibson, B.K., et~al. 2001b, in preparation
\reference Grebel, E.K., Braun, R. \& Burton, W.B. 2000, \baas, 196, \#28.09
\reference Klypin, A., Kravtsov, A.V., Valenzuela, O. \& Prada, F. 1999, \apj,
522, 82
\reference Lu, L., Sargent, W.L.W., Savage, B.D., et~al. 1998, \aj, 115, 162
\reference Majewski, S.R., Ostheimer, J.C., Kunkel, W.E., et~al. 1999, 
in New Views of the Magellanic Clouds,
ed. Y.-H. Chu et~al. (Dordrecht: Kluwer), 508
\reference Mihos, C. 2001, these proceedings
\reference Moore, B. \& Davis, M. 1994, \mnras, 270, 209
\reference Putman, M.E., Gibson, B.K., Staveley-Smith, L., et~al. 1998, 
Nature, 394, 752
\reference Richter, P., de~Boer, K.S., Widmann, H., et~al. 1999, Nature, 402,
386
\reference Robertson, J.G., Schwarz, U.J., van~Woerden, H., et~al. 1991,
\mnras, 248, 508
\reference Ryans, R.S.I., Keenan, F.P., Sembach, K.R., et~al. 1997, \mnras,
289, 83
\reference Sahu, M.S. \& Blades, J.C. 1997, \apj, 484, L125
\reference Sembach, K.R., Savage, B.D. \& Massa, D. 1991, \apj, 372, 81
\reference Tosi, M. 1988, \aap, 197, 47
\reference Tufte, S.L., et~al. 2001, in preparation
\reference van~Woerden, H., Schwarz, U.J., Peletier, R.F., et~al. 1999, Nature,
400, 138
\reference Wakker, B.P. 1990, PhD Dissertation, Groningen
\reference Wakker, B.P. \& Schwarz, U. 1991, \aap, 250, 484
\reference Wakker, B.P. \& van~Woerden, H. 1997, \araa, 35, 217
\reference Wakker, B.P., Howk, J.C., Savage, B.D., et~al. 1999, Nature, 402,
388
\reference Weiner, B.J., Vogel, S.N. \& Williams, T.B. 2001, these proceedings
\reference Wolfire, M.G., McKee, C.F., Hollenbach, D., et~al. 1995, \apj,
453, 673
\reference Yoshizawa, A. 1998, PhD Dissertation, Tohoku University
}
\end{references}
\end{document}